\documentclass{article}
\usepackage{graphicx} 
\usepackage{amsmath}
\usepackage{amssymb}
\usepackage[letterpaper, margin=1in]{geometry}
\usepackage{authblk}
\usepackage{natbib}
\usepackage{hyperref}
\usepackage{tikz}
\usetikzlibrary{arrows.meta}
\usepackage[onehalfspacing]{setspace}
\usepackage{float}

\title{From description to design: Automated engineering of complex systems with desirable emergent properties}
\author[1]{Thomas F. Varley}
\author[1,2]{Josh Bongard}

\affil[1]{Vermont Complex Systems Institute, University of Vermont, Burlington, VT, USA}
\affil[2]{Department of Computer Science, University of Vermont, Burlington, VT, USA}

\date{\today}

\begin{document}

\maketitle

\begin{abstract}
    Across the sciences, researchers have developed hundreds of statistics that describe the collective behavior of complex systems. 
    These statistics are typically used passively, to describe a system after the fact, so it is rarely learned whether the patterns captured actually matter.
    The problem of engineering systems that display desirable emergent properties is a less developed area, as the very nature of emergent phenomena challenges the typical bottom-up approach to engineering. 
    Here we propose a top-down approach, re-purposing the large number of descriptive statistics as objective functions in an optimization pipeline. 
    Since this method lets a collective property instantiated directly, it makes long-standing correlational questions amenable to experimentation.
    We demonstrate this concept on four models taken from four different areas of science: a Kuramoto coupled-oscillator model (physics), a Wilson-Cowan network of excitatory and inhibitory processes (neuroscience), an SIRS pandemic spread model (epidemiology), and an echo-state network (machine learning). 
    In each case, we take measures commonly used to describe behavior and feed those descriptive statistics back into the system to sculpt desirable macro-scale properties into the fundamental generative dynamics. 
    We find that this approach is successful across domains and systems, and the inclusion of ``emergent" measures can actually improve the ability of some systems to perform tasks. 
    We argue these threads can be unified into a general practice of emergent engineering that combines traditional task-oriented, bottom-up approaches to engineering with top-down guidance of collective, emergent properties to build resilient, capable, complex systems. 
\end{abstract}

\subsection*{Significance statement}
Scientists who study complex systems such as brains, ecosystems and societies have built a large toolkit of statistics for describing their collective behavior. 
But describing a pattern is not the same as knowing whether it matters, or being able to produce it. 
We show that these descriptive statistics can be turned around and used as design objectives: a computer adjusts a system’s individual parts until the desired collective pattern appears. 
Across four fields, the approach produced the targeted patterns, made them robust to disturbance, improved a machine-learning system’s forecasts, and found an epidemic intervention that preserved more social contact than a standard strategy. 
This offers a general way to move complexity science from description toward experiment and engineering.

\section{Introduction}
One of the fundamental purposes of modern science is to rigorously describe patterns observed in the natural world. 
The earliest works of modern science show this fascination with pattern recognition, from medieval Arab scholar Ibn al-Haytham's work on optics to the revolutionary astronomical insights of Copernicus and Kepler. 
In fields such as physics and chemistry, the formalization of patterns in Nature into mathematical models was key to the development of modern engineering capabilities. 
In other areas of science, such as biology, neuroscience, and climate science, the ability to describe systems with formal statistics has not yet led to similar breakthroughs in fine-grained control. 

Why is this? What makes biology different from rocket science?

One possible answer is that the two fields describe fundamentally different \textit{kinds} of systems. 
Cybernetician and early complexity theorist Warren Weaver introduced a now-classic taxonomy of systems \cite{weaver_science_1948}. 
The first are simple systems, which are composed of a small number of elements that interact in largely deterministic ways.
This combination of low dimensionality and predictable interactions makes them amenable to models with analytic solutions, and consequently, to control. 
Classical physics and chemistry are examples of simple systems that we understand deeply enough to engineer them. 
The second class of systems are the disorganized complex systems. 
These are systems with large numbers of elements, but trivial interactions between them, which allows the micro-scale details to be coarse-grained away and the macro-scales can be treated as analytically tractable ``simple systems" in their own right. 
Thermodynamics and fluid mechanics are two classic examples of these kinds of systems.
The final, and most interesting, class of systems are the ``organized complex systems."
These are systems with large numbers of elements that in non-trivial ways. 
These are truly complex; too high-dimensional and causally dense to be admit either analytically solvable models or simplifying coarse-grainings, these systems display interesting properties at every scale. 
Almost every aspect of biology, from cellular biology, to neuroscience, to ecology form these organized complex systems, and these are the systems that have proved most resistant to design and control. 

A fundamental problem facing any would-be engineer of organized complex systems is that, typically, the features that we would most like to control are ``emergent", macro-scale features, while we generally can only intervene on the micro-scale of individual elements.
Ground-up engineering of desirable emergent properties in complex systems is difficult-to-intractable, even for simple systems. 
Doing so would require building a mapping between what micro-scale interventions propagate up through the causal hierarchy to produce predictable changes in the emergent, macro-scale features - but as Warren Weaver pointed out, this is axiomatically difficult, due to the inherent nature of organized complex systems. 

An alternative approach might be to go ``top down" - begin with a desirable emergent, macro-scale property and somehow ``tune" the micro-scale to make the emergent property a natural attractor. 
This naturally occurs in some systems, such as those that display self-organizing criticality \cite{bak_self-organized_1991}, but in general, this is not a viable path towards a general theory of emergent engineering, as it requires some self-organizing mechanism to be already built-in to the systems micro-scale elements. 

Many different fields of engineering have run into this problem, and developed their own domain-specific approaches trying to solve it. 
In material sciences, the challenge of engineering materials with prescribed global features is a fundamental one, and inverse-design approaches \cite{torquato2009} leveraging both evolutionary \cite{miskin2013} and differentiable optimization are a promising area of modern research \cite{rocks2017,goodrich2021}.
In artificial life research, guided self-organization \cite{prokopenko2009} and morphogenesis \cite{doursat2013} are similarly foundational problems, with a long-term vision being the development of top-down pipelines for synthetic biology \cite{kriegman_scalable_2020,pezzulo2016}. 
Problems of network control theory occupy a similar domain as well \cite{liu2011,gu2015}, and in a similar space of controlling dynamics on networks is the problem of optimal synchronization design \cite{fazlyab2015}.  
Finally, and most relevant to our understanding of ``emergence" here, is the early literature on using information-theoretic descriptors to design robotic systems \cite{sporns_theoretical_2002,sporns_evolving_2006,martius2013}, although this approach has been comparatively less well-developed due to the difficulty of on-line estimation of information-theoretic measures and uncertainty about what the relevant applications are. 
It is from this final lineage that our current work takes inspiration. 
These literatures all share a common logic: specify a desired collective outcome, then explore parameter space to find the set of parameters that produce the target behavior.
Unfortunately, all of these lineages have developed largely in parallel, with limited interdisciplinary cross-talk between them. 
Furthermore, they typically focus on bespoke, domain-specific features of interest (a material property, a particular dynamical regime, etc), and what is missing is a general approach that can be applied in arbitrary domains and with arbitrary goals.
Here, we propose one such general framework that leverages the already-existing rich library of descriptive statistics and turns them into tools of design.

\begin{figure}[H]
    \centering
    \includegraphics[width=\linewidth]{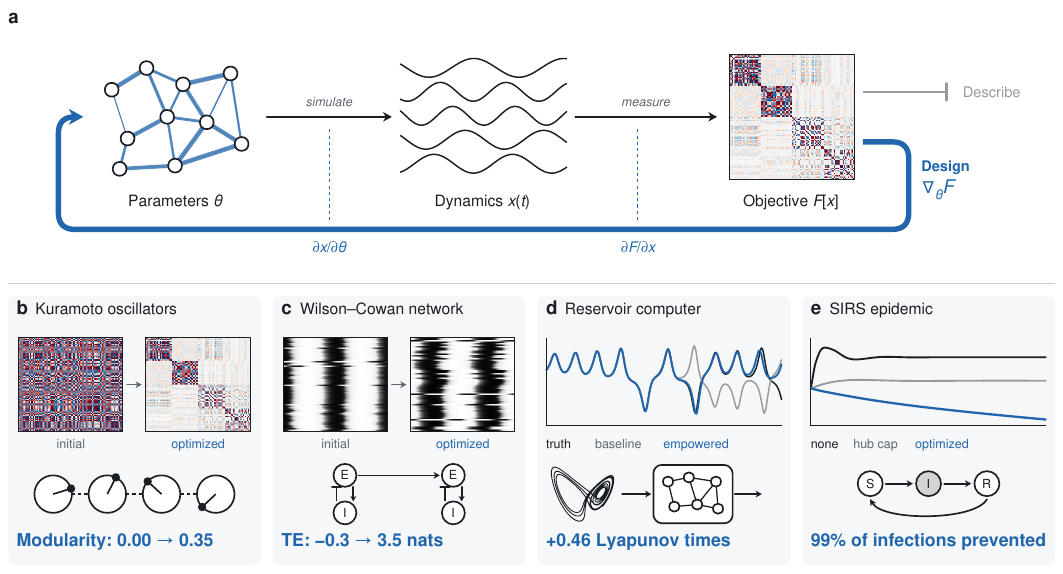}
    \caption{\textbf{A general framework for optimizing emergent features into complex, dynamical systems.}
    \textbf{a:} The parameters $\theta$ of a given system (e.g. couplings in a Kuramoto coupled-oscillator model, given in blue) are used to generate trajectories over time, which are summarized by some differentiable statistic $F[x]$ (in this example, a modified version of the modularity \cite{newman_modularity_2006} is computed for the covariance matrix). 
    In the typical, one-way pipeline, analysis ends with this descriptive statistic (gray). 
    Here, the gradient $\nabla_\theta F = \tfrac{\partial F}{\partial x}\tfrac{\partial x}{\partial \theta}$ is propagated back through the measurement and the simulation to update $\theta$ (blue), closing the loop and turning the statistic into a design target.
    \textbf{b}--\textbf{e:} Four applications, one each from physics, neuroscience, machine learning and epidemiology.
    \textbf{b:} Physics: 100 Kuramoto oscillators optimized for modularity ($Q$) of the functional connectivity network. 
    After training, $Q$ rose from an initial median value of $0.02$ to $0.37$.
    \textbf{c:} Neuroscience: 64 Wilson--Cowan populations optimized for transfer entropy from one half of the network $X$ (32 oscillators) to the other half $Y$. 
    Null-corrected $\mathrm{TE}_{X\to Y}$ rose from $-0.3$ to $3.5$ nat.
    \textbf{d:} Machine learning: an echo-state reservoir optimized for per-neuron empowerment. 
    The curves show a forecast of a chaotic Lorenz systems, from the baseline (gray) and empowered (blue) reservoirs against the true trajectory (black).
    The valid prediction time rose by $0.46$ Lyapunov times on average, beating a control in 57 of 64 seeds. 
    \textbf{e:} Epidemiology: an SIRS epidemic on a high-school contact network \cite{mastrangelo2015highschool}, optimized to reduce infections while keeping contact. Curves show disease prevalence in epidemics with no intervention (black), a targeted hub-cap intervention (gray), and the optimized contact mask (blue). 
    In the suppression regime ($\alpha = 0.5$--$0.9$), the optimized interventions prevented $98.9 \pm 0.4\%$ of infections while keeping $70\%$ of contact, $6.9$ percentage points more than hub targeting at the same level of infection.}
    \label{fig:master}
\end{figure}

Given a differentiable descriptive statistic and a differentiable system, if the computation of the summary statistic and the modeling of the system itself are all done with differentiable operations in a single computational graph, it is possible to back-propagate the gradient of the summary statistic all the way back through the data analysis pipeline, the generative process, and to the parameters that define the system itself. 
This contrasts with what might be called the ``standard" pipeline for summary statistics, where the process of description is causally ``one-way": a system is observed, data is collected, and the resulting summary/description/model has no impact on the system of origin. 
By feeding the results of the data analysis back into the generative process (see Figure \ref{fig:master}), we can create a closed loop that unifies description and design in a way that allows us to turn almost any macro-scale summary statistic into a scaffold that can guide a gradient descent optimizer towards a micro-scale configuration of parameters without requiring human-in-the-loop engineering of the microscale configurations. 

\section{Results}
In this paper we describe four systems, each taken from a different domain of modern science: networks of coupled Kuramoto oscillators (physics, \cite{kuramoto_self-entrainment_1975}), Wilson-Cowen models of excitatory/inhibatory populations (neuroscience, \cite{wilson1972}), epidemic spreading on social networks (epidemiology, \cite{pastorsatorras2001}), and reservoir computing for the prediction of chaotic systems (machine learning, \cite{surez2024}). 
See Figure \ref{fig:master}.
In each case, we identify commonly-used descriptive statistics that describe patterns of global activity and repurposed them to be differentiable objective functions in a gradient descent training set-up, demonstrating the duality between description and design.
In two systems (the reservoir computer and the epidemic spread modeling), we additionally show that informed optimization can produce systems with objectively desirable behaviors \cite{grasso_empowered_2022,10.1145/3583131.3590469} - showing that this is a promising area of practical engineering research in addition to a theoretical curiosity. 

\subsection{Information-theoretic auxiliaries improve performance in a reservoir computer}

\begin{figure}[]
  \centering
  \includegraphics[width=\textwidth]{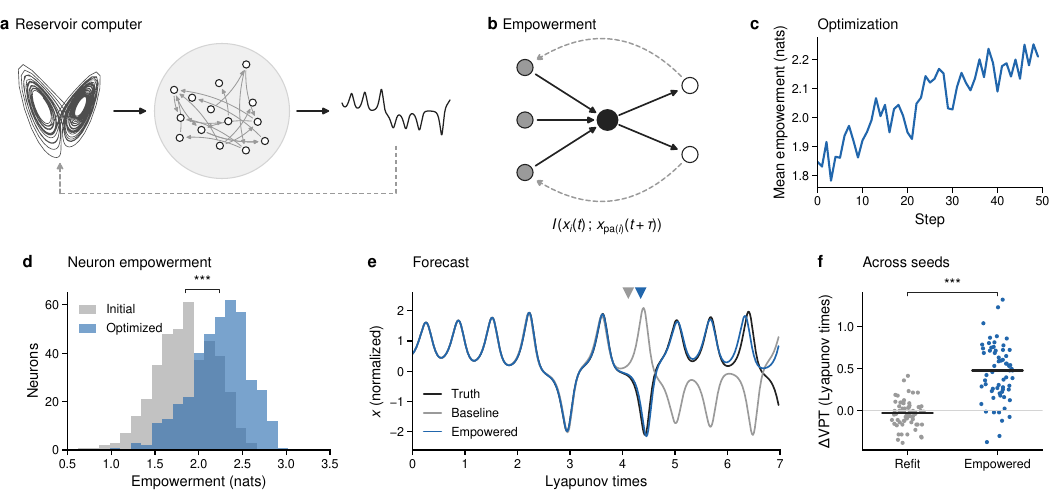}
    \caption{\textbf{Maximizing per-neuron empowerment in a reservoir computer extends chaotic forecasts.}
  \textbf{a:} An echo-state network ($N = 400$ leaky-tanh units, sparse directed recurrent
  weights at 5\% density) is driven a chaotic Lorenz system \cite{DeterministicNonperiodicFlow} and is trained to forecast future states.
  The recurrent, input and bias weights stay fixed, only the per-unit leak rates, per-unit input
  gains, and the global recurrent gain are learned.
  \textbf{b:} The empowerment of neuron $i$ (black) is the Gaussian mutual information between its
  present state $x_i(t)$ and the future states of its parents in the directed graph (gray),
  $x_{\mathrm{pa}(i)}(t+\tau)$, with $\tau = 5$~steps. 
  Because the graph is recurrent, $i$'s activity
  returns to its own parents (dashed), closing an action--perception loop.
  \textbf{c:} Mean empowerment over all neurons during gradient ascent (50 epochs),
  after which the readout is refitted.
  \textbf{d:} Distribution of per-neuron empowerment before (gray) and after (blue) optimization.
  The distributions differ (two-sample Kolmogorov-Smirnov test, $N = 400$ neurons,
  $p = 8.8\times10^{-35}$).
  \textbf{e:} An example held-out forecast from the baseline (gray) and empowered
  (blue) reservoirs against the true trajectory (black). 
  Triangles mark the valid prediction time (VPT) when the normalized error first exceeds 0.4.
  \textbf{f:} Change in VPT relative to baseline for 64 independently seeded reservoirs (Lyapunov times; bars, medians), after maximizing empowerment (empowered) or, as a control, after only refitting the readout on fresh data (refit). 
  Median difference $+0.50$; two-sided Wilcoxon signed-rank test, $p = 2.3\times10^{-11}$.}
  \label{fig:empowerment-explainer}
\end{figure}

Recent work \cite{grasso_empowered_2022,10.1145/3583131.3590469,tolle_evolving_2024} suggests that optimizing information-theoretic functions can improve the performance of artificial machine-learning and artificial intelligence systems.
Grasso \& Bongard showed that including an auxiliary information-theoretic measure of action/sensor control called empowerment \cite{klyubin2005empowerment} improved the performance of neural cellular automata trained on a morphological shape-matching task.
Similarly, Tolle et al.\ showed that optimizing a measure of higher-order, synergistic temporal information \cite{rosas_reconciling_2020} enhanced the ability of an echo-state network to predict the future state of a chaotic dynamical system.
Both approaches, however, relied on discrete estimators of mutual information that are not amenable to gradient-based optimization, and instead required thousands of generations of sample-intensive evolutionary optimization.

Here, we present a synthesis of Grasso et al and Tolle et al., showing that including neuron-level empowerment as an auxiliary function can improve the performance of an echo-state network using a fully differentiable, gradient-based optimization pipeline.
An echo-state network with a reservoir of 400 leaky-tanh neurons, structured into a sparse (5\%) directed random network, was driven by the three-dimensional Lorenz system \cite{DeterministicNonperiodicFlow}.
The readout weights were fitted by ridge regression to predict the next state of the system (Fig.~\ref{fig:empowerment-explainer}a).
The empowerment of neuron $i$ was the mutual information $I(X_{i}(t) ; X_{\mathrm{Pa}(i)}(t+\tau))$, where $\mathrm{Pa}(i)$ is the set of presynaptic parents of $X_i$ and $\tau = 5$ steps (Fig.~\ref{fig:empowerment-explainer}b).
Mutual information was computed with a Gaussian covariance estimator, and the average empowerment over all 400 neurons was maximized by gradient ascent for 50 steps (Fig.~\ref{fig:empowerment-explainer}c).
We compared the valid prediction time (VPT) \cite{pathak2018,vlachas2020} of echo-state networks trained to forecast the Lorenz system with and without empowerment-optimized reservoirs.

Fifty epochs of training were enough to significantly increase average neuron empowerment across 64 trials, from $1.88 \pm 0.11$ nat before shaping to $2.21 \pm 0.07$ nat after (Kolmogorov--Smirnov test, $p < 10^{-33}$; Fig.~\ref{fig:empowerment-explainer}d).
For each reservoir, we measured VPT relative to its own pre-shaping baseline ($\Delta$VPT).
As a control, we refitted the readout of the unshaped reservoir on fresh data.
Refitting alone left VPT unchanged ($\Delta$VPT $= -0.03 \pm 0.16$ Lyapunov times), whereas empowerment shaping increased it ($+0.46 \pm 0.34$; Fig.~\ref{fig:empowerment-explainer}f).
Empowered reservoirs outperformed their refit controls in 57/64 seeds (paired Wilcoxon test, $p = 2.3\times10^{-11}$).

These results provide a compelling proof-of-concept that direct-task optimization can be efficiently enhanced by preceding task training with an abstract, domain-general objective, replicating \cite{grasso_empowered_2022,10.1145/3583131.3590469} with efficient, differentiable estimators. 
In doing so, it shows that the benefits of engineering emergent properties may not be merely of theoretical interest, but actually useful in applied contexts as well. 

\subsection{Building global structure into networks of coupled oscillators}
\begin{figure}[]
  \centering
  \includegraphics[width=\textwidth]{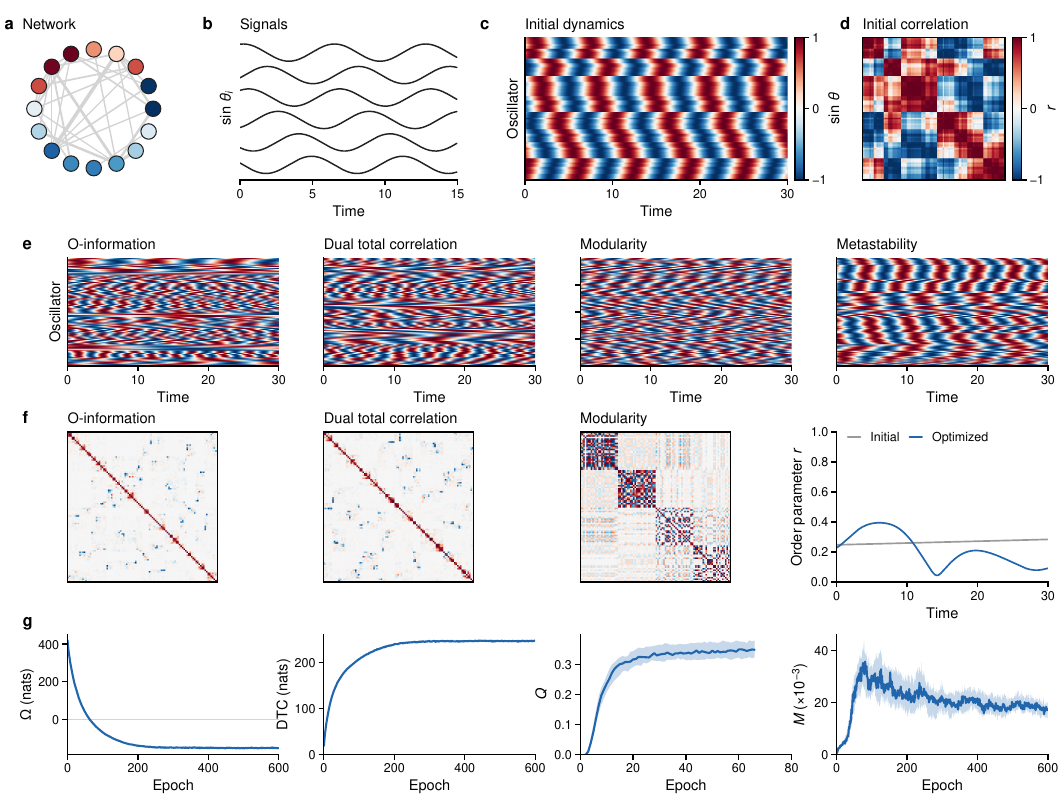}
    \caption{\textbf{Gradient-based design of Kuramoto networks for higher-order information structure.}
  \textbf{a:} A network of phase oscillators shown schematically.
  Node color shows $\sin\theta_i$ at one moment in time.
  \textbf{b:} The system is integrated from a batch of random initial phases, and then phases are sine-transformed. 
  Time series for six oscillators of the initial, un-optimized network are shown.
  \textbf{c:} A carpet plot showing $\sin\theta_i(t)$ for all oscillators from an un-optimized network. 
  Note how the system naturally falls into global patterns of redundant synchrony.
  \textbf{d:} Correlation matrix $\Sigma$ of the $z$-scored signals shown in (e).
  The correlation matrix is what each information-theoretic objective is computed from using Gaussian estimators.
  \textbf{e:} Carpet plots for systems optimized for: negative O-information (synergy-dominance), dual total correlation, modularity, and metastability.
  \textbf{f:} Correlation matrices after optimization, as in (d). 
  For metastability, the order parameter $r(t)$ of the initial (gray) and optimized (blue) networks.
  \textbf{g:} The trajectory of training the objective: median (line) and interquartile range (band) over 25 independent runs. }
  \label{fig:kuramoto-explainer}
\end{figure}

Collective behavior in coupled oscillators, from synchronization to chimera and metastable states, is a canonical example of emergence in physics \cite{kuramoto_self-entrainment_1975,shanahan2010metastable}.
We tested whether gradient descent could build prescribed global structure into a system of 100 Kuramoto oscillators by optimizing both the adjacency matrix $A$ and the intrinsic frequencies $\omega$ (Fig.~\ref{fig:kuramoto-explainer}a,b).
We optimized four targets in separate runs: two information-theoretic measures of higher-order structure: the O-information (a heuristic measure of redundancy/synergy balance) \cite{rosas_characterising_2025} and the dual total correlation (a measure of total information-sharing) \cite{abdallah_measure_2012}, as well as the metastability \cite{shanahan2010metastable}, and a modularity-like measure sensitive to the contrast between within-module and between-module edges in the associated functional connectivity network \cite{sporns2016} (Fig.~\ref{fig:kuramoto-explainer}c).

The initial networks were dominated by coherent traveling waves and strong redundancy ($\Omega = 422 \pm 1$ nat).
Minimizing $\Omega$ drove the networks through zero into the synergy-dominated regime ($\Omega = -153 \pm 5$~nat), and maximizing DTC raised it from $18.2 \pm 0.4$ to $247 \pm 3$~nat.
Both effects survived a shift-surrogate null that removes all dependencies between oscillators while preserving each oscillator's own dynamics: the null-corrected O-information fell from $192 \pm 2$ to $-92 \pm 5$~nat, and the null-corrected DTC rose from $-41 \pm 1$ to $116 \pm 3$~nat (25/25 runs each).
Maximizing modularity raised $Q$ from $0.019 \pm 0.004$ to $0.37 \pm 0.04$, and maximizing metastability raised $M$ fivefold, from $(0.6 \pm 0.3)\times10^{-3}$ to $(3.0 \pm 0.9)\times10^{-3}$.

To test whether the optimized properties are attractors of the learned dynamics rather than artifacts of particular initial conditions, we perturbed each optimized network with white phase noise ($\sigma = 1$ rad) and let it evolve for the same duration.
All four properties recovered: $\Omega = -151.6 \pm 10.1$~nat after perturbation (before, $-150.4 \pm 9.4$), DTC $= 246.2 \pm 4.5$~nat ($247.7 \pm 4.7$), $Q = 0.37 \pm 0.04$ ($0.37 \pm 0.04$) and $M = (3.1 \pm 1.2)\times10^{-3}$ ($(2.8 \pm 1.4)\times10^{-3}$; paired Wilcoxon tests, all $p > 0.2$; Supplementary Fig.~\ref{fig:si-perturbation}a).

In the echo-state network, we were optimizing a global property of the system by enhancing the predictive power of individual elements.
Here, we demonstrate that it is possible to sculpt truly higher-order, collective properties such as a synergy and redundancy at their native macro-scale, and that these properties become genuinely robust features of the system itself (robust to perturbation). 
While the Kuramoto model does not have a natural task for which it can be trained the way the echo-state network does, future work on training systems of coupled oscillators may discover similar effective auxiliary functions \cite{wang2024}.

\subsection{Building time-directed computational dynamics into biological models}

\begin{figure}[]
  \centering
  \includegraphics[width=\textwidth]{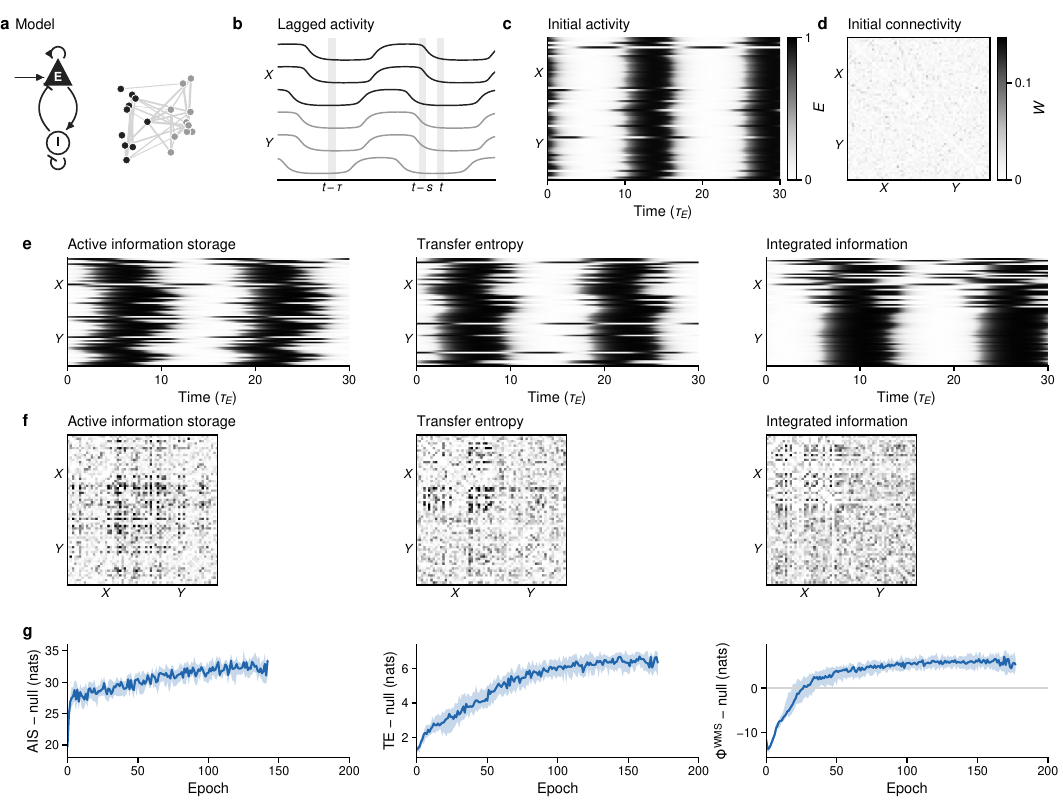}
    \caption{\textbf{Optimizing temporal information structure in a Wilson--Cowan network.}
  \textbf{a},~Each of $N = 64$ nodes is an excitatory--inhibitory population pair with local couplings $w_{EE}$, $w_{EI}$, $w_{IE}$, $w_{II}$ and external drive $P$ (arrows are excitatory; bars are inhibitory). 
  Nodes are coupled through excitatory long-range connections $W$ and the network is split into two halves, $X$ (32 nodes, black) and $Y$ (32 nodes, gray). 
  The weights, external drive, and the four local weights are trained by gradient descent.
  \textbf{b:} Traces are excitatory activity of three nodes from each half of the TE-optimized network shown below.
  \textbf{c:} Carpet plot of excitatory activity $E$ of all 64 nodes of the initial network.
  \textbf{d:} Initial long-range connectivity $W$.
  \textbf{e:} Carpet plots for networks optimized for each of the information dynamics (AIS, TE, $\Phi^{WMS}$. 
  \textbf{f:} Long-range connectivity after optimization, as in (d).
  \textbf{g:} The corrected measures over training; median (line) and interquartile range (band) over 25 independent runs, each stopped at convergence
  after 50--232~epochs.
  Raw measures and nulls are shown in Supplementary Fig.~\ref{fig:si-wilson-surrogate}.}
  \label{fig:wilson-cowan-explainer}
\end{figure}

The dynamics of complex systems are often analogized to computation, and complex systems scientists have developed statistical proxies for the basic primitives of computation. 
Systems are described as storing, transferring and integrating information over time, and information-theoretic measures can quantify these patterns in quantitative data \cite{lizier_jidt_2014,bossomaier_introduction_2016}.
We asked whether these time-directed forms of information processing could be built into a biophysical model of cortical population dynamics: a network of 64 coupled Wilson-Cowan excitatory/inhibitory populations \cite{wilson1972}.
Since information dynamics have been observed at multiple scales (such as between groups of thousands of neurons), rather than optimize element-level information dynamics, we partitioned the system into  halves ($X$ and $Y$), each containing 32 elements and optimized macro-scale information dynamics across the partition.
The learned parameters were the long-range coupling $W$, the external drive $P$ and the four local coupling weights learned (Fig.~\ref{fig:wilson-cowan-explainer}a).

In separate runs we maximized three measures of how the network's past informs its present (Fig.~\ref{fig:wilson-cowan-explainer}b): the active information storage of the whole system (AIS), the transfer entropy from $X$ to $Y$ ($\mathrm{TE}_{X\to Y}$) \cite{schreiber_measuring_2000}, and the whole-minus-sum integrated information $\Phi^{\mathrm{WMS}}$ \cite{balduzzi_integrated_2008,mediano_integrated_2022}.
For all measures, the past was embedded at two lags, $t-20$ steps, and $t-300$ steps, so that the objective sees both the recent and the more distant past and accounts for both when optimizing.
To avoid spurious dependencies caused by autocorrelation, we corrected each measure by subtracting off a shift-surrogate null, ensuring that the optimizer was building genuine information dynamics rather than exploiting lower-order features. 

All three corrected measures increased over training (Fig.~\ref{fig:wilson-cowan-explainer}d).
Against a held-out null that was never optimized, the excess AIS rose from $4.0 \pm 7.0$ to $16.5 \pm 9.3$~nat (paired Wilcoxon test, $p = 5\times10^{-6}$), from $-0.3 \pm 1.7$ to $3.5 \pm 2.5$~nat for $\mathrm{TE}_{X\to Y}$ ($p = 3\times10^{-6}$), and from $-3.8 \pm 2.4$ to $4.9 \pm 3.1$~nat for $\Phi^{\mathrm{WMS}}$ ($p = 6\times10^{-8}$).
$\Phi^{\mathrm{WMS}}$ successfully crossed the boundary from global redundancy into genuine information integration.
The optimized transfer was directed: $\mathrm{TE}_{X\to Y}$ exceeded $\mathrm{TE}_{Y\to X}$ in 88\% of the runs ($10.4 \pm 1.9$ vs $8.1 \pm 2.2$~nat; $p = 1.5\times10^{-6}$), compared with 44\% before optimization.
Across runs, the long-range coupling became more heterogeneous, and in the example networks the near-synchronous oscillation of the initial network gave way to staggered activity.

\begin{figure}[H]
  \centering
  \includegraphics[width=\textwidth]{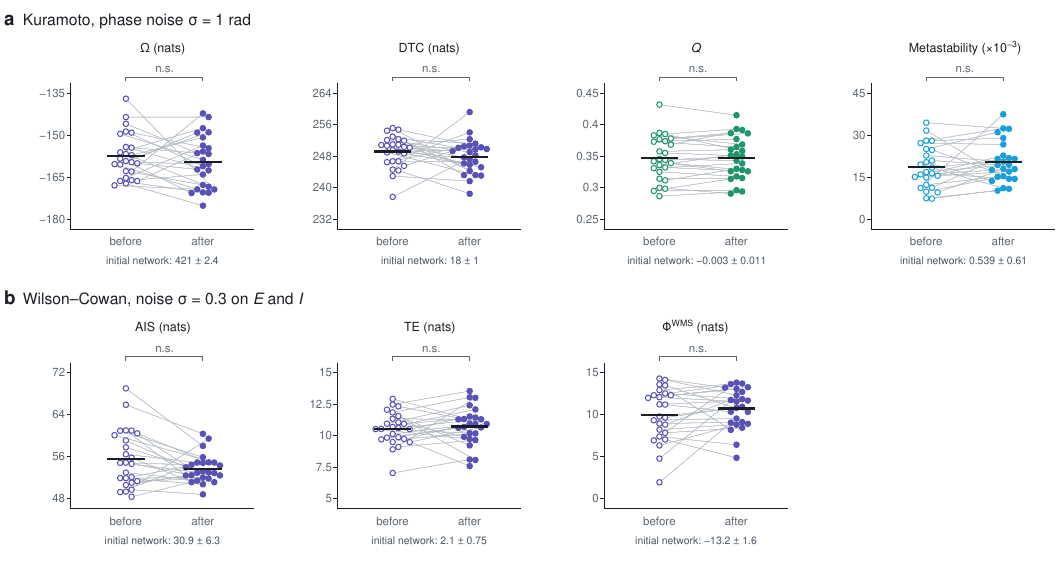}
  \caption{\textbf{Optimized properties are recovered after perturbation.}
  Tracking the effects of injecting white noise into a system on the target measures to compare pre- and post-perturbation values. 
  \textbf{a:} Kuramoto networks optimized for $\Omega$ (minimized), DTC, modularity $Q$ and
  metastability, with phase noise $\mathcal{N}(0, 1)$~rad. 
  \textbf{b:} Wilson--Cowan networks
  optimized for AIS, $\mathrm{TE}_{X\to Y}$ and $\Phi^{\mathrm{WMS}}$, with noise
  $\mathcal{N}(0, 0.3^2)$ added to $E$ and $I$. 
  No measure differs between
  before and after (paired Wilcoxon tests, all $p > 0.05$; n.s.).}
  \label{fig:si-perturbation}
\end{figure}

The optimized properties were again attractors of the learned dynamics.
After white noise ($\sigma = 0.3$) was added to the excitatory and inhibitory activity, $\mathrm{TE}_{X\to Y}$ ($10.7 \pm 1.5$~nat after, $10.5 \pm 1.3$ before) and $\Phi^{\mathrm{WMS}}$ ($10.7 \pm 2.4$ and $10.0 \pm 3.2$~nat) fully recovered, and AIS recovered to $53.7 \pm 2.7$~nat from $55.6 \pm 5.5$ (all $p > 0.05$), far above the initial networks ($30.9 \pm 6.3$~nat; Supplementary Fig.~\ref{fig:si-perturbation}b).

\subsection{Optimizing global properties autonomously discovers effective pandemic mitigation strategies}

\begin{figure}[]
  \centering
  \includegraphics[width=\textwidth]{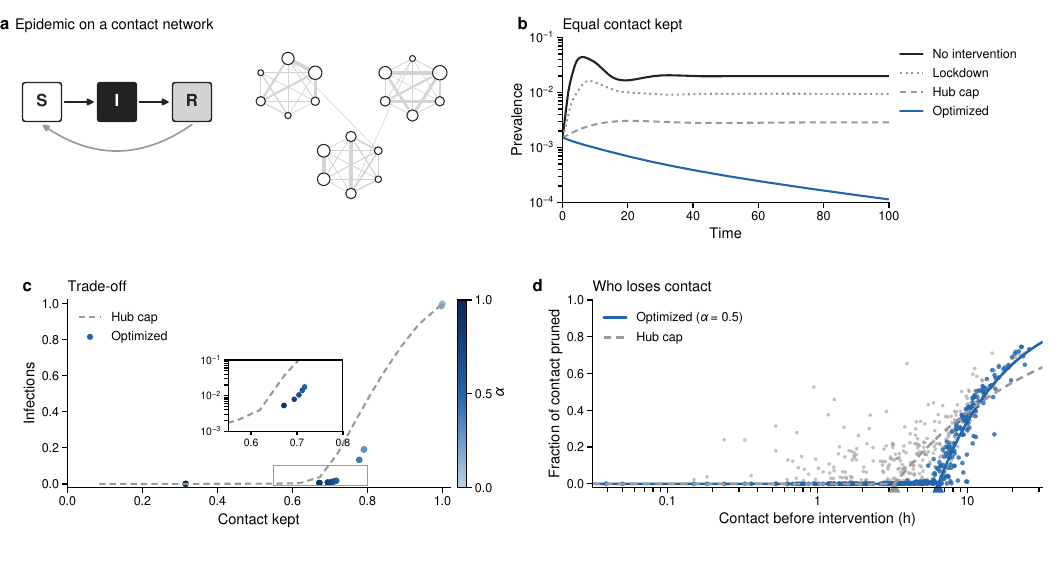}
  \caption{\textbf{Designing epidemic interventions on a real social contact network.}
  \textbf{a:} An epidemic modeled by an SIRS system spreads over the contact network of a French high school \cite{mastrangelo2015highschool}. 
  The graph shows the six most social students in three different classes, with node size and edge weight tracking each student's total contact hours.
  The learned intervention is to weaken existing contacts (decreasing weights of edges), and the ultimate goal is to minimize the total amount of edge mass removed (i.e. minimizing disruption to the social network) while simultaneously minimizing the cumulative incidence of infections. 
  \textbf{b:} Disease prevalence after outbreaks with no intervention (black), and with three
  interventions that keep the same total contact (71.6\%): a uniform lockdown (dotted), a targeted intervention that that only modified the highest-degree nodes
  (the hub-cap intervention dashed) and the optimized intervention (blue).
  \textbf{c:} Infections versus contact kept for optimized interventions (11 values of
  $\alpha$, 3 replications each), compared to the targeted hub-cap intervention.
  At matched infections, the optimized mask keeps $+0.07$ more contact than the hub cap in both the mitigation ($\alpha = 0.3-0.4$) and suppression ($\alpha = 0.5-0.9$)
  regimes. 
  Inset: the suppression regime on a logarithmic infections axis. 
  \textbf{d:} Fraction of each student's total contact reduced, plotted against their pre-intervention total contact (edge strength), for
  the optimized intervention (blue) and the hub cap reaching the same infections (gray).
  Lines, fits of the threshold form $\max\!\left(0,\, 1 - (d_c/d)^{b}\right)$, for which an exact cap on
  each student's contact at $d_c$ gives $b = 1$.
  The optimized curve best fit found $d_c = 6.4$~h,
  $b = 0.92$, $R^2 = 0.93$ while the hub cap found $d_c = 3.3$~h, $b = 0.44$, $R^2 = 0.54$.
  The optimized intervention finds a threshold similar to the the hub-cap, however it is more targeted in its edge removals: almost all students below the threshold see no change in their social contacts.
  In contrast, the hub-cap null modifies the contacts of many students below the threshold because those students interact with targeted high-degree nodes.}
  \label{fig:sirs-explainer}
\end{figure}

Our final experiment in automated design came from epidemiology: could an optimizer find minimally invasive alterations to a social network that also limited the spread of a contagion?
This is a generally tricky trade-off, as connectivity in a social network is a key facilitator of epidemic spread. 
Solving this problem, then, requires clever targeting of specific parts of the network.
We ran a susceptible-infected-recovered-susceptible (SIRS) model on a real-world network (the SocioPatterns high-school contact network \cite{mastrangelo2015highschool}).
The network comprises 327 students in 9 classes, and the weight of the edge between two students is the number of 20-s intervals in which they were in face-to-face proximity over five days.
Outbreaks were seeded in five random students and the model was integrated over 100 time units.
Epidemic severity was measured as the cumulative incidence ($CI$); the integral of the mean per-person rate of new infections, normalized by its value on the unmodified network (See Materials and Methods for details).
Disruption to the network was measured as the fraction of edge-level contact kept; $\Delta A = \sum_{ij}A_{ij}/\sum_{ij}A^{0}_{ij}$, where $A^{0}$ is the weighted adjacency matrix of the original contact network.
The loss $\alpha\cdot CI-(1-\alpha)\cdot \Delta A$ was minimized, reducing infections while preserving social contact (Fig.~\ref{fig:sirs-explainer}c).
The gradient was backpropagated to a mask $m$ that scales the original connectivity matrix, $A = m \odot A^{0}$ with $0 < m_{ij} < 1$; the optimizer could weaken existing edges but it could not create new.

We compared the optimized interventions with one based on targeting hubs, which are known to drive epidemic spreading \cite{pastorsatorras2001}.
The hub cap scales down the contacts of every student whose total contact exceeds a fixed threshold, which also weakens their ties to students below it.
This is analogous to preferentially vaccinating high-contact individuals such as nurses and doctors \cite{nunner2022}.

We ran three replicates for each of eleven values of $\alpha \in \{0, 0.1, \ldots, 1\}$ and compared the contact kept and infections of each optimized intervention with the hub cap reaching the same level of infection (Fig.~\ref{fig:sirs-explainer}d).
For $\alpha \le 0.2$, intervening was not worthwhile and the network was left almost intact ($99.9 \pm 0.1\%$ of contact kept).
In the mitigation regime ($\alpha = 0.3$--$0.4$), the optimized interventions reduced infections to $16.2 \pm 3.2\%$ of the unmitigated epidemic while keeping $78.4 \pm 0.7\%$ of contact, against $71.7 \pm 1.0\%$ for the hub cap.
In the suppression regime ($\alpha = 0.5$--$0.9$), they reduced infections to $1.1 \pm 0.4\%$ while keeping $70.0 \pm 1.6\%$ of contact, against $63.0 \pm 0.7\%$ for the hub cap.
The optimized interventions kept more contact than the hub cap in all 21 runs across both regimes, by $6.9 \pm 0.9$ percentage points.
Only with $\alpha = 1$, where contact carries no cost, did the optimum fall below the hub cap ($31.5 \pm 0.0\%$ vs $35.1 \pm 0.2\%$).
We also explored two alternative comparisons, one which targeted edges for removal based on edge betweenness and one based on spectral graph theory. 
Neither performed as well as the hub-cap intervention, for visualization see Supplementary Figure \ref{fig:si-nulls}

Did the optimizer develop an interpretable strategy?
Without being told which structural features matter, it independently rediscovered hub targeting (Fig.~\ref{fig:sirs-explainer}d).
Students with less than a critical threshold of about 6.4 hours of cumulative contact lost only $0.5 \pm 1.6\%$ of their contact, whereas those above it lost $33.1 \pm 20.5\%$, with pruning rising steeply with degree, close to an exact per-student cap.
The hub cap, in contrast, removed $13.3 \pm 11.2\%$ of the contact of students that were already below this threshold, because capping a hub also cuts its ties to less social students (since less social students are likely to share edges with hubs).
The optimizer avoided this by pruning hub-hub connections: if we (somewhat arbitrarily) classify the 30\% most-social students as hubs, we find that the optimizer kept 91\% of contact between hubs and less social students but only 39\% of contact among hubs.
This strategy appears to be an emergent attack on the ``rich club" structure of social networks \cite{vandenheuvel2011,huang2021}, preferentially disrupting interactions between two high-degree nodes rather than high-low or low-low dyads. 

This final set of results serves as validation for the idea that multi-objective optimization can learn causally significant patterns in complex systems without any kind of pre-trained knowledge.
Not only did it successfully discover the salience of high-degree nodes, it was able to leverage that implicit knowledge to beat the raw, degree-capping intervention that an actor with knowledge of network science might construct. 

\section{Discussion}
In this paper, we showed how gradient-based optimization pipelines can be used to automate the design of complex systems with specified emergent properties. 
By recasting descriptive summary statistics as loss functions, we can take advantage of the large library of descriptive measures scientists have developed over the decades; transforming them from post-hoc descriptions to usable design constraints.
We took a high-level approach, exploring a broad sample of possible applications, including optimizing global properties, meso-scale patterns, and showing how abstract auxiliary functions (e.g. empowerment \cite{grasso_empowered_2022,10.1145/3583131.3590469}) can improve performance above straight task-oriented training. 
This is a very small collection of possible research directions: the space of descriptive statistics that could be repurposed is vast: in information theory alone, there are many different measures of ``emergence" that could be explored \cite{rosas_reconciling_2020,klein_emergence_2020,barnett_dynamical_2021,varley_emergence_2022}, and other measures from graph theory, dynamical systems theory, and computer science could be similarly adapted. 
Since many measures of ``emergence" hinge on some notion of the past state of a system disclosing predictive information about its own future state \cite{varley_flickering_2023}, extending the time-lagged objectives optimized here (AIS, TE and $\Phi^{\mathrm{WMS}}$) will be an important next step. 

This work will hopefully make it possible to critically assess what, if any, insights scientists have actually gained from developing this plethora of descriptive statistics. 
Since high-level, emergent features are generally impossible to intervene upon, relationships between descriptive statistics and other features are typically correlational; the question of the causal efficacy of properties like synergy and metastability is a significant outstanding question.
Scientists may be able to show that, for example, synergistic information changes with loss of consciousness \cite{luppi_synergistic_2024}, or over the course of infant development \cite{varley_emergence_2025}, but without being able to manipulate these features, any attempt to make sense of them is inherently an exercise in \textit{post-hoc} storytelling. 
Does recognizing patterns of higher-order information flow in the brain actually give us any fundamental, actionable insight? 
Or are we just engaged in a kind of statistical stamp-collecting for its own sake? 
By turning descriptive statistics into loss functions and designing them into systems from scratch, we can begin to transition the field of complex systems science from one based on correlational and descriptive methods to a more interventional one. 
If one wants to know if synergistic information fusion is important for multi-modal sensory integration in a system, change the amount of synergy and see what happens. 
While we may never be able to do this in living, biological systems, the march of simulation technology towards ever-greater detail and veracity may soon make it possible to rigorously study the effects of controlling emergence in naturalistic systems. 

This approach bears similarities to the paradigm of ``emergent engineering" introduced by Ulieru and Doursat \cite{ulieru2011}, who described a novel engineering paradigm emerging in the post-cybernetic era in which engineers serve as ``enablers" who guide self-organizing, cybernetic systems towards desirable global configurations, rather than ``dictators" who prescribe the entire structure via a totalizing blueprint. 
Our approach operationalizes the paradigm that Ulieru and Doursat specified at the theoretical level. 
\cite{ulieru2011} sets out general principles, but gives no formal way to specify the desired global behavior, and crucially, no general procedure for building emergent behavior from micro-scale rules.
Using differentiable programming and optimization on descriptive summary statistics, our approach resolves both ambiguities and, as a bonus, also automates the role of the ``enabler." 
Rather than trying to find local rules that produce collective behavior (e.g. finding the specific rules that implement a parity gate in cellular automata \cite{betel2012}), our approach to ``emergent engineering" instead asks what collective behavior best facilitates some particular set of sub-goals and steers the self-organizing system into those global configurations automatically.
One important difference remains, however: their original proposal focuses on systems that continue to flexibly adapt through time, whereas our pipeline designs parameters offline. 
Closing this loop and allowing online adaptation towards the global target is a natural extension for follow-up research.

Variations on this theme have been explored in specific contexts (such as work on ``intrinsic motivations" in artificial life such as empowerment \cite{klyubin2005empowerment} and free-energy minimization \cite{badcock2022}), but this approach likely has more general applications to fields outside of specific academic disciplines. 
Increasingly, the challenges facing humanity are being recognized as challenges in controlling and modeling complex systems \cite{varley_information_2025}, and in such a world, the ability to act on emergent, collective dynamics will be a vitally important skill. 

\subsubsection*{Limitations}
This approach has a number of limitations to be addressed in future work. 
The largest is the reliance on Gaussian estimators for efficient, differentiable quantification of the various summary statistics, which restricts us to parametric measures. 
Since complex systems typically display nonlinearities, this is a potentially significant limitation. 
For some measures, such as information-theoretic predictive power, non-parametric, neural network-based estimators have been proposed \cite{karl2017,mohamed2015}, but it reamins to be seen if they can be efficiently included in this kind of pipeline. 
More generally, differentiability is a high bar for many measures of collective organization: descriptive measures that are naturally discrete (such as topological data analysis \cite{chazal2021} or state-transition dynamics \cite{klein_emergence_2020}) will require adaptation or an alternative approach \cite{mathieucarriere2021}. 
Even in naturally continuous systems such as chaotic dynamical systems, gradients computed over long rollouts in chaotic systems can display pathological behavior \cite{mikhaeil2022,metz2021} which may require further refinement or regularization than what is demonstrated here. 
Finally, this approach assumes that an aspiring systems engineer has access to the correct ``ground-truth" forward model. 
In some fields, the causal logic of foundational models have been extremely well-validated (such as SIRS models), while in other areas (such as biology) significant discrepancies and simplifying assumptions complicate this approach. 
We do not feel that any of these limitations fundamentally compromise the basic premise of our approach: instead, they represent practical challenges that can be overcome by future researchers refining this approach. 

\section{Conclusions}
In this paper, we introduced a general pipeline for using gradient descent to automate the process of designing high-dimensional systems with desirable emergent properties. 
The key contribution is the recognition that the extensive library of descriptive statistics developed over the years by scientists from many fields can be repurposed into a set of loss functions that can be optimized on, with the gradient descent algorithm doing the fine tuning that would be intractable by hand. 
Ultimately, this work aims to help scientists working on complex systems transition from describing the systems under study towards being able to engineer and control them.
By automating one of the key bottlenecks in the design of complex systems (the difficulty of mapping micro-scale features to global emergent properties), we hope to facilitate new advances in multiple fields of science and engineering. 

\subsection*{Author Contributions}
TFV and JB conceptualized the research project. TFV wrote the software, performed experiments, analyzed data, and wrote the initial draft of the manuscript. JB provided feedback and supervision throughout. 

\subsection*{Conflict of interest disclosure}
The authors declare no conflicts of interest. 

\subsection*{Funding}
The authors received no special funding for this work. 

\subsection*{Data and code availability}
The raw results will be included as supplementary information.
Code is available at \url{https://www.github.com/thosvarley/description2design}

\bibliography{automated_design}

\section*{Supplementary Information}

\section*{Methods}

\subsection*{General approach}
Each system is a set of ordinary differential equations $\dot{x} = f_\theta(x)$, with parameters $\theta$, which are optimized by gradient descent targeting a scalar descriptor $F[x]$ of the simulated
trajectories. 
Systems were integrated with a fixed-step, fourth-order Runge--Kutta integrator, and gradients $\nabla_\theta F$ were computed from the adjoint sensitivity method using \texttt{torchdiffeq} \cite{chen2018neural} in PyTorch. 
Each run batched eight simulations, each initialized with independent random initial conditions, but sharing parameters $\theta$. 
Loss was averaged over the batch.
Parameters were optimized with Adam \cite{kingma_adam_2017}. 
After every update, coupling matrices were symmetrized, clipped at zero (where necessary), and given a
zero diagonal. 
Unless stated otherwise, 25 independent runs were performed for each objective.

\paragraph{Information-theoretic estimators.} All information quantities were computed using Gaussian estimators \cite{varley_information_2025} from covariance matrices of $z$-scored signals and with a ridge term added to the diagonal. 
The total correlation was estimated by:
\begin{align}
    \mathrm{TC} = -\frac{\log|\mathbf{R}|}{2}
\end{align}
with correlation matrix $\mathbf{R}$.

The dual total correlation was estimated by:
\begin{align}
\mathrm{DTC} = \frac{\log\!\big(|\mathbf{R}|\prod_i (\mathbf{R}^{-1})_{ii}\big)}{2}; 
\end{align}
The O-information \cite{rosas_quantifying_2019} was estimated by: 
\begin{align}
\Omega = \mathrm{TC} - \mathrm{DTC}
\end{align}
Finally, mutual informations between sets $A$ and $B$ were estimated by: 
\begin{align}
I(A;B) = \tfrac{1}{2}(\log|\Sigma_A| + \log|\Sigma_B| - \log|\Sigma_{AB}|)
\end{align}
and conditional mutual informations analogously. 

\paragraph{Shift-surrogate nulls.} Information-theoretic estimates computed on autocorrelated trajectories are biased upwards, and autocorrelated dynamics alone can produce spurious values. 
Each measure $F$ was therefore compared with a null $F_{\text{null}}$ generated with a circular-shift model.
The time series for each element was shifted by a random offset, which disrupts higher-order dependencies while preserving each elements first-order properties.
Nulls were averaged over four shifts and are differentiable, so the corrected measure
$F - F_{\text{null}}$ could itself be optimized. 

\subsection*{Kuramoto oscillators}
A network of $N = 100$ coupled oscillators oscillators evolved according to the standard rule \cite{kuramoto_self-entrainment_1975}:

\[\dot\theta_i = \omega_i + {N}\sum_j A_{ij}\sin(\theta_j - \theta_i)\] 

\begin{figure}[H]
  \centering
  \includegraphics[width=\textwidth]{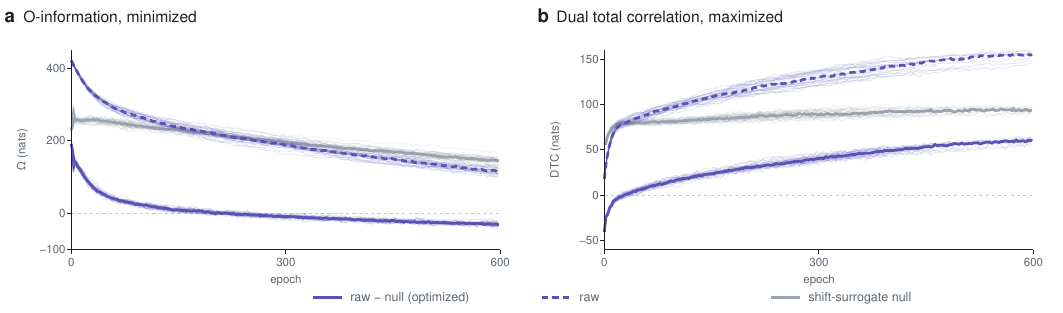}
  \caption{\textbf{Kuramoto networks optimized on null-corrected objectives.}
  Training trajectories for corrected information measures: the raw value minus its shift-surrogate null. 
  The null was generated by shifting every oscillator by an independent random offset (four surrogates per epoch), which removes all dependencies between oscillators while preserving each oscillator's own first-order dynamics (autocorrelationm etc).  
  \textbf{a:} O-information, minimized: the corrected value falls from $151 \pm 4$ to $-25 \pm 5$~nat
  (mean $\pm$ SD). 
  \textbf{b:} Dual total correlation,
  maximized: $-24 \pm 1$ to $42 \pm 6$~nat. 
  Both move in the intended direction in all runs (paired Wilcoxon tests, $p = 6\times10^{-8}$), but the O-information runs remain redundancy-dominated in raw terms (final $\Omega = 119 \pm 14$~nat), falling below their null ($144 \pm 11$~nat) rather than below zero. 
  Networks optimized on the raw objectives (main text) show larger null-corrected effects when
  re-simulated from fresh initial conditions, so they are the ones reported.}
  \label{fig:si-kuramoto-surrogate}
\end{figure}

The coupling matrix was initialized as $A = (B + B^\top)/2$ with
$B_{ij} \sim \mathrm{Exp}(1)/100$ and zero diagonal, and all natural frequencies as $\omega_i = 1$.
The optimizer could modify $A$ and $\omega$. 
Initial phases were drawn uniformly on $[0, 2\pi)$ and the system was integrated over $t \in [0, 150]$ in 3000 steps. 
Descriptors were computed from the correlation matrix $\mathbf{R}$ of the signals $\sin\theta_i(t)$ (ridge $10^{-4}$). 
We also maximized the metastability
$M = \mathrm{Var}_t\, r(t)$, the temporal variance of the order parameter
$r(t) = |N^{-1}\sum_j e^{i\theta_j(t)}|$ \cite{shanahan2010metastable}, computed after discarding the
first third of each trajectory. 
The reported runs optimized the raw
objectives, but to test $\Omega$ and DTC against chance, the initial and optimized parameters of each run
were re-simulated from four fresh initial conditions, and each value was compared with its
shift-surrogate null (described above). 
As a control, $\Omega$ and DTC were also optimized in their
shift-surrogate-corrected forms, $\Omega - \Omega_{\text{null}}$ and
$\mathrm{DTC} - \mathrm{DTC}_{\text{null}}$ (25 runs each), which gave smaller corrected effects
(Supplementary Fig. \ref{fig:si-kuramoto-surrogate}).
All objectives used a learning rate of $10^{-2}$ and at most 600 epochs, with early stopping if the the
standard deviation of the last 50 losses fell below 1 ($\Omega$, DTC) or 0.01 ($Q$).

\subsection*{Wilson--Cowan network}
Each of $N = 64$ nodes comprised excitatory and inhibitory populations \cite{wilson1972}:
\begin{align*}
\tau_E \dot{E}_i &= -E_i + S\big(w_{EE}E_i - w_{EI}I_i + P_i + \kappa\textstyle\sum_j W_{ij}E_j\big),\\
\tau_I \dot{I}_i &= -I_i + S\big(w_{IE}E_i - w_{II}I_i\big),
\end{align*}
with $S(x) = [1 + e^{-a(x-\vartheta)}]^{-1}$, $a = 1.2$, $\vartheta = 2.8$, $\tau_E = 1$,
$\tau_I = 5$ and $\kappa = 1$. 
The long-range coupling $W$ was initialized as the Kuramoto $A$,
with $P_i = 2.5$ and $(w_{EE}, w_{EI}, w_{IE}, w_{II}) = (10, 12, 9, 3)$.
The optimizer was allowed to modify $W$, $P$ and the four
local weights. 
Initial states were drawn uniformly on $[0, 1]$, the system was integrated over $t \in [0, 150]$ in 3000 steps, and the first half was discarded. 
From the $z$-scored excitatory activity $Z = (X, Y)$, with $X$ the first 32 nodes and $Y$ the rest, we embedded the past at two lags, $Z^{-} = (Z_{t-s}, Z_{t-\tau})$ with $s = 20$ steps ($\tau_E$) and $\tau = 300$ steps ($15\,\tau_E$), to compute the active information storage: \[\mathrm{AIS} = I(Z_t; Z^{-})\] the transfer entropy \cite{bossomaier_introduction_2016, schreiber_measuring_2000}:  \[\mathrm{TE}_{X\to Y} = I(Y_t; X^{-} \mid Y^{-})\] and the whole-minus-sum integrated information \cite{balduzzi_integrated_2008}:  \[\Phi^{\mathrm{WMS}} = I(Z_t; Z^{-}) - I(X_t; X^{-}) - I(Y_t; Y^{-}).\]

Each measure $F$ was maximized separately in its shift-surrogate-corrected form, $F - F_{\text{null}}$ and averaged over the batch. 
Optimization used a learning rate of
$10^{-2}$, halved every 25 epochs, gradient-norm clipping at 1, and stopped when the absolute slope
of a linear fit to the last 50 losses fell below $10^{-3}$~nat per epoch (at most 600 epochs).

\subsection*{Reservoir computer}
An echo-state network \cite{jaeger2001} of $N = 400$ units evolved according to the standard rule:

\begin{align} 
x_{t+1} = (1 - a) \odot x_t + a \odot \tanh\!\big(g_{\mathrm{rec}} W x_t + g_{\mathrm{in}} \odot W_{\mathrm{in}} u_t + b\big)
\end{align}

with a readout:
\begin{align}
y_t = W_{\mathrm{out}}\,[x_t, x_t^{2}, 1]
\end{align}. 
The weight matrix $W$ had Gaussian entries at 5\% density,
rescaled to spectral radius 1.05. 
Entries of $W_{\mathrm{in}}$ and $b$ were drawn uniformly from
$[-0.1, 0.1]$. 
$W$, $W_{\mathrm{in}}$ and $b$ were fixed. 
The learned parameters were the per-unit leak rates $a_i = 0.02 + 0.97\,\sigma(\ell_i)$ (initially 0.5), per-unit input gains $g_{\mathrm{in},i}$ and a global recurrent gain $g_{\mathrm{rec}} \in [0.1, 1.4]$ (both initially 1).
The input was the chaotic Lorenz system \cite{DeterministicNonperiodicFlow} ($\sigma = 10$, $\rho = 28$, $\beta = 8/3$), integrated with
$\Delta t = 0.02$ after a 4000-step transient and normalized per coordinate. 
The readout was fitted by ridge regression (regularization $10^{-2}$) on 48 teacher-forced windows of 400 steps following a 150-step washout.

The empowerment of neuron $i$ computed as:
\begin{align}
\mathrm{Emp}_i = I\big(x_i(t);\, x_{\mathrm{pa}(i)}(t + \tau)\big)
\end{align}
with $\tau = 5$ steps.
$\mathrm{pa}(i)$ are node $i$'s presynaptic parents
\cite{klyubin2005empowerment}.
Mean empowerment over neurons was maximized for 50 Adam steps (learning rate $3\times10^{-3}$, gradient-norm clipping at 1) on batches of 32 teacher-forced windows of 300 steps, after which the readout was refitted. 
Forecast skill was quantified by the valid prediction time (VPT): the time at which the root-mean-square error of an autonomous closed-loop forecast first exceeded 0.4 \cite{pathak2018,vlachas2020}, expressed in Lyapunov
times ($\lambda_{\max} = 0.9056$). 
We compared 64 independently seeded reservoirs before and after shaping, with a control in which the
readout was refitted on fresh data without shaping.

\subsection*{SIRS epidemic}
We simulated a susceptible-infected-recovered-susceptible epidemic spread model on the network of high-school social contacts given by the SocioPatterns study \cite{mastrangelo2015highschool}.
The network consists of 327 students, broken down into 9 classes. 
The adjaceny matrix $A^{0}_{ij}$ the number of 20-s intervals in which students $i$ and $j$ were in
face-to-face proximity over five days.
The network was rescaled so that $\rho(A^{0}) = 1$. 
The epidemic followed the standard form:

\begin{align}
\dot{S}_i &= -\beta S_i \sum_j A_{ij}I_j + \delta R_i,\\
\dot{I}_i &= \beta S_i \sum_j A_{ij}I_j - \gamma I_i,\\
\dot{R}_i &= \gamma I_i - \delta R_i
\end{align}

with $\gamma = 1$, $\delta = 0.1$ and $\beta = R_0/\rho(A^{0})$, and $R_0 = 3$. 
Each epoch seeded 8 outbreaks in 5 random students ($I_i = 0.1$), integrated over $t \in [0, 100]$ in 2000 steps. 
The optimizer's goal was to learn to remove edge mass in such a way as to balance two conflicting impulses: minimize edge mass removal (preserving the initial social network), while simultaneously minimizing infections: $\alpha\cdot\text{infections} - (1-\alpha)\cdot\text{contact kept}$. 
Infections were the cumulative incidence of infections, normalized by the average number of infections that occurred in a parallel pandemic with the same initial conditions but no mitigation intervention. 
Contact kept was $\sum_{ij}A_{ij}/\sum_{ij}A^{0}_{ij}$. 
We swept values of $\alpha$ from $\{0, 0.1, \dots, 1\}$, with 3 replicates per $\alpha$ (600 epochs, learning rate
$5\times10^{-2}$). 
Final interventions were evaluated on 64 held-out outbreaks.
We compared the optimized results against an alternative intervention that targeted high-degree nodes (analogous to vaccinating healthcare workers in a pandemic setting \cite{nunner2022}). 
The hub cap sets a limit on each student's total contact time and scales down the total edge mass of every student above above the threshold until they hit the limit - essentially uniformly weakening the weights of all connections incident on a hub. 
We swept the limit in increments of 5 from the 5th to the 100th percentile. 

\subsection*{Perturbation test}
To test whether optimized properties are attractors of the learned
dynamics rather than properties of particular initial conditions, each optimized network (and its
initial network) was simulated from a random initial condition, white noise was added to its final
state (Kuramoto: phases, $\sigma \in \{0.5, 1, 2\}$~rad; Wilson--Cowan: $E$ and $I$,
$\sigma \in \{0.1, 0.3\}$, clipped to $[0, 1]$), and the network was simulated again for the same
duration. Each measure was computed as in training on both segments and compared across the 25 runs.

\subsection*{Statistics}
Paired changes were tested with two-sided Wilcoxon signed-rank tests and distributions with
two-sample Kolmogorov--Smirnov tests. Dynamics shown in
the figures were re-simulated from saved parameters.

\newpage 
\section*{Supplementary Results}

\begin{figure}[H]
  \centering
  \includegraphics[width=\textwidth]{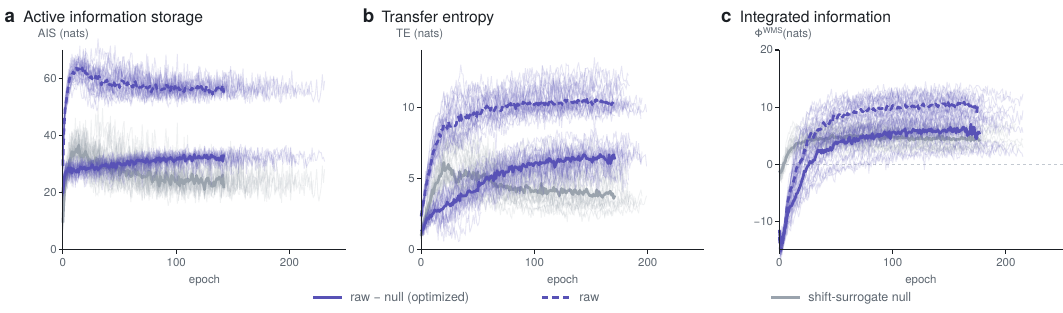}
  \caption{\textbf{Raw measures and shift-surrogate nulls during Wilson--Cowan optimization.}
  Training trajectories of the runs in Fig. \ref{fig:wilson-cowan-explainer}d. 
  Three different runs are plotted: the raw objective (dashed purple), the circle-shifted null (grey), and the corrected objective: raw - shifted null (dashed purple).
  \textbf{a:} Active information storage. \textbf{b:} Transfer entropy
  $\mathrm{TE}_{X\to Y}$. 
  \textbf{c:} Integrated information $\Phi^{\mathrm{WMS}}$.}
  \label{fig:si-wilson-surrogate}
\end{figure}

\begin{figure}
    \centering
    \includegraphics[width=0.8\linewidth]{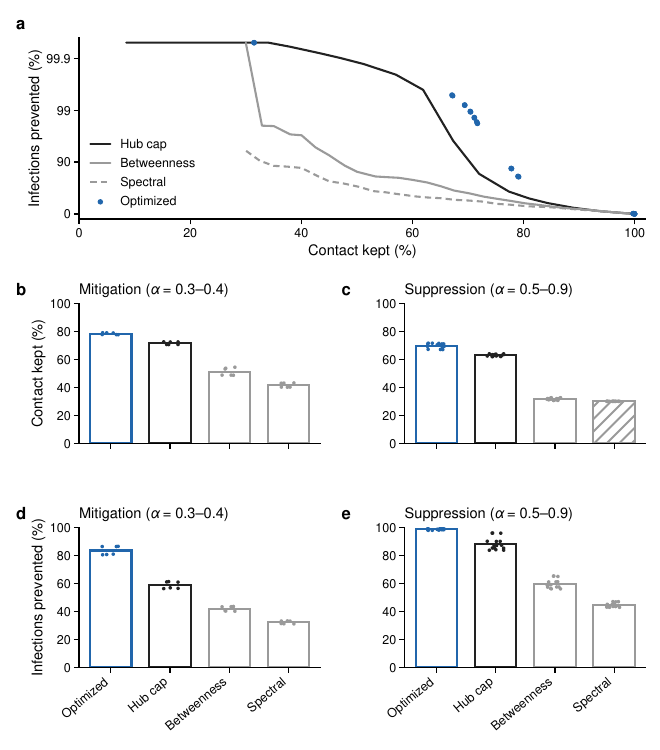}
      \caption{\textbf{Optimized interventions keep more contact than targeted intervention. a:}
      Here we compared the optimized intervention against three interventions that explicitly leverage graph-theoretic knowledge. 
      The hub-cap model targets high-degree nodes specifically (analogous to the public health intervention of vaccinating healthcare workers \cite{nunner2022,pastorsatorras2001}).
      Betweenness removes contacts in order of their weighted edge betweenness (the fraction of shortest paths that pass through a given edge, with path length defined as the inverse of contact time \cite{girvan2002, brandes2001, holme2002}. 
      Spectral removes edges in order of how each edge contributes to the epidemic threshold, which is set by the largest eigenvalue of the connnectivity matrix \cite{chakrabarti2008}. Removing the
      contact between students $i$ and $j$ lowers $\lambda$ by approximately $2 w_{ij} u_i u_j$, where $w_{ij}$ is their contact time and $u$ the eigenvector of $\lambda$ \cite{vanmieghem2011}.
      \textbf{a:} The Pareto front of contacts kept (as a percentage) against infections presented.
      \textbf{b-c:} Contact kept at fixed infections. 
      For each optimized run, infections are held at the level reached by the optimizer, and each baseline's contact kept at that level is read from its curve in \textbf{a}.
      \textbf{d-e:} Infections prevented at fixed contact. For each optimized run, contact kept is held at the level that run kept, and each baseline's infections prevented at that level are read from its curve in \textbf{a}.
      Left, mitigation ($\alpha = 0.3$--$0.4$, 6 runs); right, suppression ($\alpha = 0.5$--$0.9$, 15 runs). 
      The spectral curve does not reach the suppression level of infection, so its bar is hatched. 
      In all runs, all three comparator interventions kept less contact (at fixed infections), and prevented fewer infections (at fixed contact-loss), than the optimized intervention.}
    \label{fig:si-nulls}
\end{figure}

\end{document}